\documentclass[apj,twocolumn,twocolappendix,numberedappendix]{openjournal}
\usepackage{newtxtext,newtxmath,enumitem,multirow,ctable}
\usepackage[T1]{fontenc}
\usepackage[breaklinks,colorlinks,citecolor=blue,urlcolor=blue]{hyperref}

\newcommand{\Alf}{{Alfv\'en}}

\newcommand{\gizmourl}{\href{http://www.tapir.caltech.edu/~phopkins/Site/GIZMO.html}{\url{http://www.tapir.caltech.edu/~phopkins/Site/GIZMO.html}}}

\newcommand{\orcidauthor}[3]{\author{\href{http://orcid.org/#1}{#2$^{#3}$}}}

\shorttitle{Lagrangian-Eulerian Magnetized Disks}
\shortauthors{Tomar \&\ Hopkins}

\begin{document}

\title{\vspace{-0.8cm}Lagrangian versus Eulerian Methods for Toroidally-Magnetized Isothermal Disks\vspace{-1.5cm}}

\author{Yashvardhan Tomar$^{1}$}
\orcidauthor{0000-0003-3729-1684}{Philip F. Hopkins}{1*}
%\orcidauthor{0000-0002-3680-5420}{Minghao Guo}{2}
%\orcidauthor{0000-0001-8479-962X}{Jonathan Squire}{3}
\affiliation{$^{1}$TAPIR, Mailcode 350-17, California Institute of Technology, Pasadena, CA 91125, USA}
%\affiliation{$^{2}$Department of Astrophysical Sciences, Princeton University, Princeton, NJ 08544, USA}
%\affiliation{$^{3}$Physics Department, University of Otago, 730 Cumberland St., Dunedin 9016, New Zealand}

\thanks{$^*$E-mail: \href{mailto:phopkins@caltech.edu}{phopkins@caltech.edu}},

\begin{abstract}
A number of simulations have seen the emergence of strongly-toroidally-magnetized accretion disks from interstellar medium inflows. Recently, \citet{guo:2025.idealized.sphere.collapse.sims.hypermagnetized.disks.resolution.dependent.on.resolving.thermal.scale.height.but.limited.physics} (G25) studied an idealized test problem of toroidally-magnetized disks in isothermal ideal MHD with an Eulerian static-mesh method, and argued the midplane behavior changes qualitatively (with a significant loss of toroidal magnetic flux) when the the thermal scale-length is resolved ($\Delta x < H_{\rm thermal}$). We rerun the G25 test problem with two Lagrangian methods: meshless finite-mass, and meshless finite-volume. We show that Lagrangian methods reproduce the high-resolution ($\Delta x \ll H_{\rm thermal}$) Eulerian G25 results. At low resolution ($\Delta x \gg H_{\rm thermal}$), behaviors differ: Lagrangian methods still lose flux and evolve ``as close as possible'' to the converged solution, while Eulerian methods show no evolution. We argue this difference in convergence behavior is related to the ability of Lagrangian codes to follow flows to an arbitrarily thin midplane layer, analogous to the well-studied difference in Jeans fragmentation problems. This and results from other higher-resolution simulations and different codes suggest that the sustained midplane toroidal fields seen in recent Lagrangian multi-scale, multi-physics simulations cannot be a numerical resolution effect, and some physical difference between those simulations and the G25 test problem explains their different behaviors.
\end{abstract}


\maketitle

\section{Introduction}
\label{sec:intro}

A number of studies have argued that accretion disks around AGN in particular could be strongly magnetized, with primarily toroidal magnetic fields and midplane magnetic $\beta_{\rm th} \equiv P_{\rm gas,\,thermal} / P_{\rm B} \ll 1$ \citep{gaburov:2012.public.moving.mesh.code,guo:2024.fluxfrozen.disks.lowmdot.ellipticals,shi:2024.seed.to.smbh.case.study.subcluster.merging.pairing.fluxfrozen.disk,hopkins:superzoom.disk,hopkins:superzoom.overview,hopkins:superzoom.imf,hopkins:superzoom.agn.disks.to.isco.with.gizmo.rad.thermochemical.properties.nlte.multiphase.resolution.studies,kaaz:2024.hamr.forged.fire.zoom.to.grmhd.magnetized.disks,wang:2025.hypermagnetized.circumbinary.disk.flux.frozen.cavity.to.pc.scales}. 
In all of these simulations, the disks emerge from attempts to more self-consistently model the accretion process from large-to-small scales with inflows from interstellar medium-type scales and structures. 
This is essential to understand, as it could radically change our understanding of the accretion disk structure, support for phases like the broad-line region and Comptonizing layers, line-driven winds, and many other properties and puzzles of AGN observations \citep[references above and][]{hopkins:multiphase.mag.dom.disks}. 

In order to better understand these disks, \citet{squire:2024.mri.shearing.box.strongly.magnetized.different.beta.states} (S25) and \citet{guo:2025.idealized.sphere.collapse.sims.hypermagnetized.disks.resolution.dependent.on.resolving.thermal.scale.height.but.limited.physics} (G25) performed some idealized experiments, both using static Cartesian grids in ATHENA, with isothermal, ideal MHD in an analytic Keplerian potential and no other physics. S25 considered a shearing-box, with different initial toroidal field strengths, while G25 considered the collapse of a homogeneous, uniformly rotating and uniformly magnetized sphere. Both found that when their spatial resolution was too poor -- specifically when $\Delta x \gtrsim H_{\rm thermal} = c_{s} / \Omega$ in terms of the thermal scale-length $H_{\rm thermal}$, isothermal sound speed $c_{s}$, and Keplerian angular frequency $\Omega = \sqrt{GM/R^{3}} = v_{K} / R$ -- the initial magnetization of the midplane was maintained indefinitely, however it was set. But when G25 increased their resolution to $\Delta x < H_{\rm thermal}$, they eventually (after hundreds of dynamical times) saw a ``vertical collapse'' of the midplane to a dense, sharply-peaked layer with midplane $\beta \sim 1$ (though a very thick layer with $\beta \ll 1$, containing an order-unity fraction of the total mass and much of the total accretion rate through the disk, still persisted indefinitely at all resolution levels). S25 showed consistent results, and how this related to the shutdown of a quasi Parker dynamo --- the mechanism by which magnetic buoyancy, turbulence, and shear drives Parker instability modes to collectively transport but also partially re-generate midplane magnetic flux --- (as originally hypothesized in \citealt{johansen.levin:2008.high.mdot.magnetized.disks}). 

However the static, Cartesian meshes used for these tests are numerically quite different from many of the original motivating simulations which use Lagrangian or quasi-Lagrangian numerical methods. And it is known that various collapse test problems often exhibit qualitatively different behavior between Eulerian and Lagrangian methods (see \S~\ref{sec:lagrangian}). We therefore repeat the experiment in G25 using multiple Lagrangian methods, to inform our interpretation of toroidally-magnetized disks and numerical convergence criteria for their behavior. \S~\ref{sec:methods} describes the problem setup and tests. \S~\ref{sec:lagrangian} discusses how this relates to other known differences between Lagrangian and Eulerian methods, and \S~\ref{sec:discuss} discusses the implications for the multi-physics simulations above. We summarize in \S~\ref{sec:summary}.

\begin{figure}
	\centering
	\includegraphics[width=0.98\columnwidth]{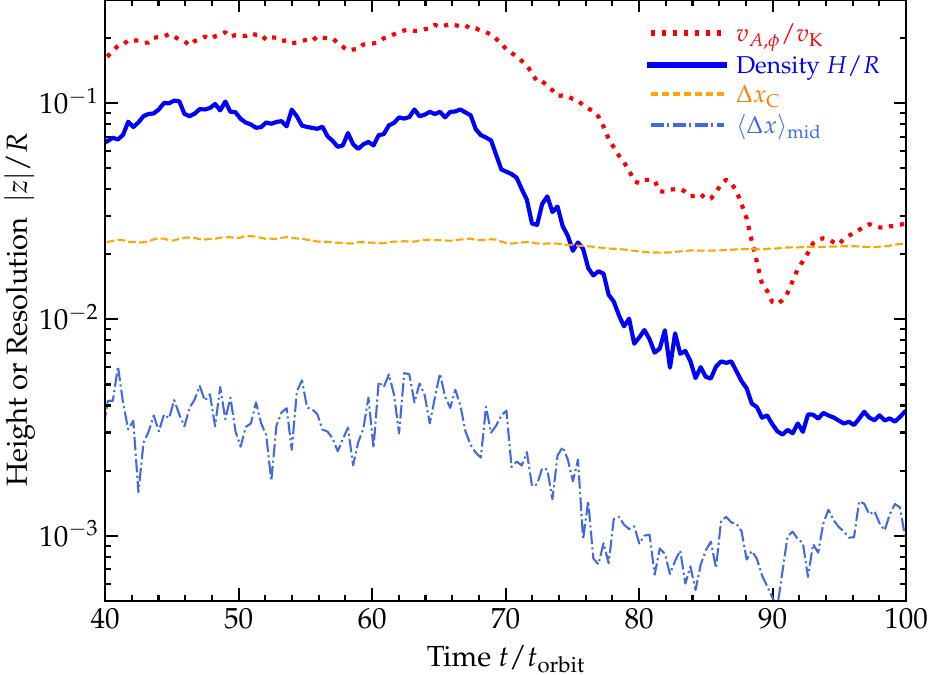}
	\vspace{-0.1cm}
	\caption{G25 collapsing disk test problem -- time evolution (\S~\ref{sec:methods}).
	We present a Lagrangian MFM simulation with \textit{unresolved} thermal scale-height ($h=10^{-4}$) and extremely low $\beta$. All quantities are measured in a narrow cylindrical annulus centered on the circularization radius $R_{\rm circ}$.
	\textit{Top panel} shows the time evolution of the Alfven scale-height $H_{B} = v_{A,\,\phi}/\Omega$ (where $\Omega$ is the orbital frequency) and the midplane density scale-height $H_{\rho}$, both normalized by radius. We also show the Cartesian-equivalent resolution $\langle\Delta x\rangle^{\rm C} = ({\rm Volume}/N)^{1/3}$ (measured over the full cylindrical volume) and the median Lagrangian resolution near the midplane $\langle\Delta x\rangle^{\rm mid}$ (measuring actual cell spacing in the dense midplane layer), both in units of $R$. The time axis extends from initial disk formation until ``collapse'' as defined by G25 occurs.
	\label{fig:timeseries}}
\end{figure}

\begin{figure}
	\centering
	\includegraphics[width=0.98\columnwidth]{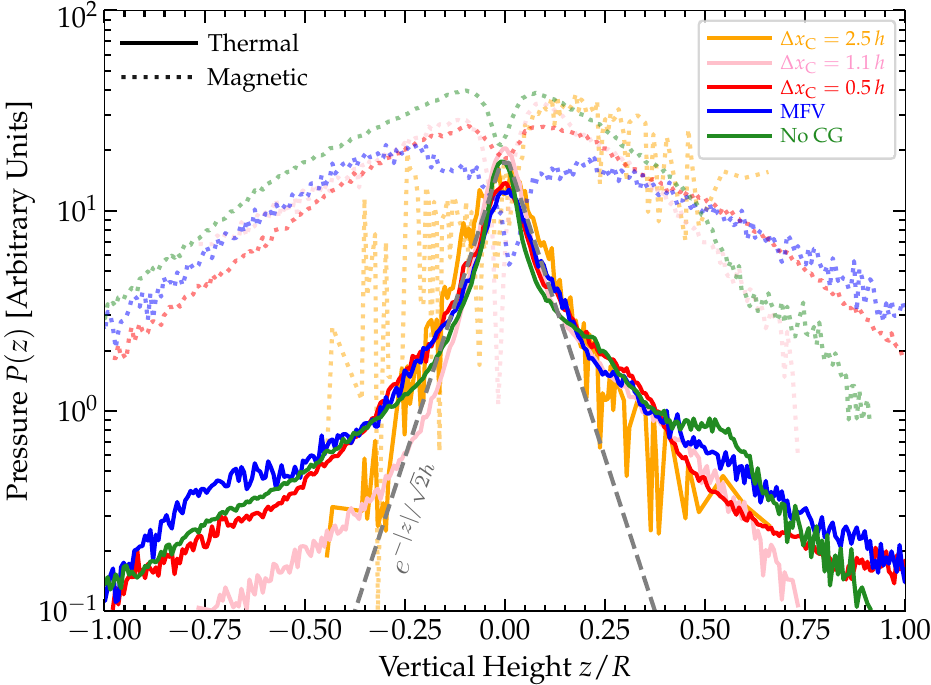}
	\includegraphics[width=0.98\columnwidth]{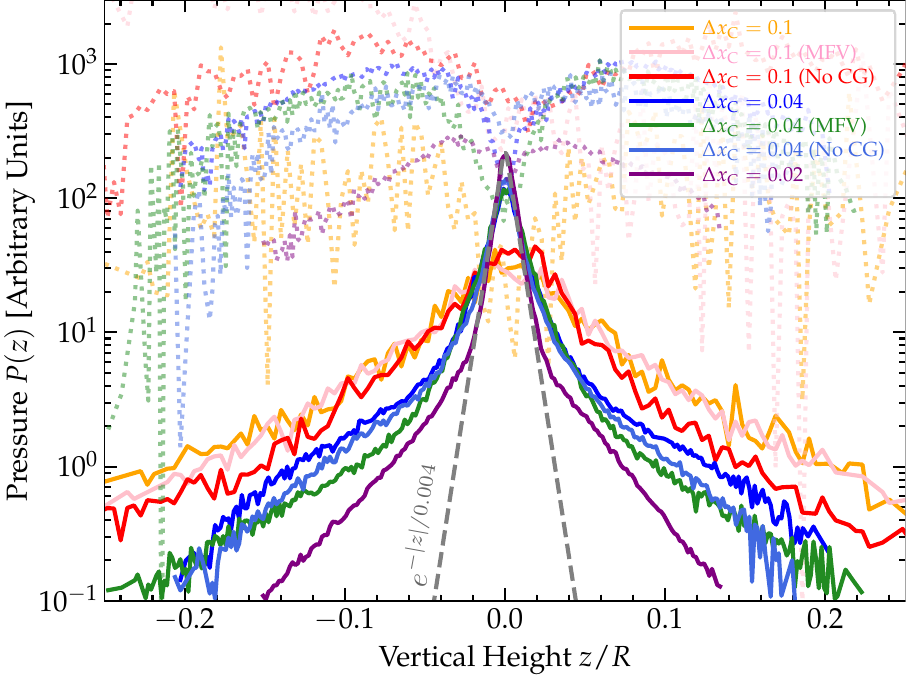}
	\vspace{-0.1cm}
	\caption{G25 collapsing disk test problem -- vertical pressure profiles (\S~\ref{sec:methods}).
	\textit{Top panel:} Vertical thermal ($P_{\rm th} = \rho\,c_{s}^{2}$) and magnetic ($P_{\rm B} = B_{\phi}^{2}/2$) pressure profiles for the thermal scale-height case ($h=0.05$) with Lagrangian MFM at three resolution levels, including one MFV run and one run without constrained-gradient MHD. Results are shown at the time when the density scale-height reaches its minimum value (marked as the ``collapse'' state in Fig.~\ref{fig:timeseries}). We overlay an exponential profile $\rho \propto \exp{(-|z|/\sqrt{2}\,h)}$ for comparison. At high resolution, converged solutions show midplane $\beta\sim1$ with density scale-height $\sim \sqrt{2}\,h$.
	\textit{Bottom panel:} Same as top, but for the unresolved case with $h=10^{-4}$. The behavior is qualitatively similar to the ``high-resolution'' ($\langle\Delta x\rangle^{\rm C} < h$) Eulerian tests in G25: the density scale-height decreases over tens of orbits, $\rho(z)$ develops a sharp peak at the midplane, and $P_{B}$ shows a characteristic minimum around $|z|\sim 0$. The slight asymmetry visible in some profiles arises from spiral structures and overdense rings that develop during collapse (see \S~\ref{sec:methods}), which create local variations in magnetic field structure at different azimuthal angles. Notably, with Lagrangian methods, collapse initiates even when $\langle\Delta x\rangle^{\rm C} \gg h$, and proceeds until the disk reaches approximately one-cell thickness, approaching the resolution limit.
	\label{fig:profiles}}
\end{figure}

\section{Methods \&\ Tests}
\label{sec:methods}

We initialize the test problem from G25, to which we refer for details. Briefly, we evolve the equations of ideal MHD without self-gravity in an analytic Keplerian potential with $G=M=1$, with an inner outflow/sink boundary at $r_{0}=1$ and initial uniform gas density $\rho_{0}=1$, velocity ${\bf v}_{0} = v_{\phi,0} \hat{\phi}$, $v_{\phi,0} = \sqrt{G M R_{\rm circ}}/R$ with $R_{\rm circ} = 16\,r_{0}$ and $R$ the cylindrical radius, and ${\bf B}_{0} = B_{\phi,0}\,\hat{\phi}$, $B_{\phi,0} = 0.0045$, outside of an initial ``tapering'' region (at $R < 8\,R_{0}$) defined in G25. The equation-of-state is set to be strictly locally-isothermal with $h \equiv c_{s}/v_{K} = c_{s}\,\sqrt{R/GM}= 0.05$, where $h = H_{\rm thermal}/R$ by definition (with $H_{\rm thermal} = c_{s}/\Omega$ the thermal scale-height), with a minimum $c_{s,\,{\rm min}} \approx 0.003$. The outer boundary is open but we only initialize gas out to a radius where the free-fall time is long compared to the time to which we evolve the simulations, so it has no effect. 
We implement this in the code GIZMO \citep{hopkins:gizmo},\footnote{A public version of the code \citep{hopkins:gizmo.public.release} is available at \gizmourl.} using the second-order reconstruction and HLLD solver. By ``strictly locally-isothermal,'' we mean that the code does not solve the full energy equation; rather, it enforces a fixed temperature $T(R)$ that varies with orbital radius. GIZMO is a multi-method code and we consider several different numerical methods: our default setup uses the meshless finite-mass (MFM) method for consistency with \citet{hopkins:superzoom.disk}, but we also consider meshless finite-volume (MFV) and (low-resolution-only) fixed Cartesian-mesh simulations. 
We also consider MFM with both the constrained-gradient CG-MHD \citep{hopkins:cg.mhd.gizmo} reconstruction method as used in \citet{hopkins:superzoom.disk}, or without CG-MHD (in which case we employ the standard Dedner divergence-suppression scheme) as in \citet{hopkins:mhd.gizmo}. We analyze the gas around $R_{\rm circ}$, as G25: defining ``collapse'' by the evolution of $H_{\rho}/R$, the density scale-height (defined as the median mass-weighted $|z|$ in a narrow annulus around $R_{\rm circ}$) and $H_{B}/R$, the \Alf\ scale-height defined by $H_{B} \equiv v_{A,\,\phi} / \Omega = |B_{\phi}|/\sqrt{\rho}\,\Omega$ where $B_{\phi}$ is measured in a small cylindrical ring with vertical extent $-H_{\rho}/2 < z < H_{\rho}/2$.  The thermal scale-height $H_{\rm thermal}/R = c_{s}/v_{K} \equiv h = 0.05$, by construction.
We also consider a test where $H_{\rm thermal}$ is set much smaller (intentionally unresolveable), multiplying $c_{s}$ and therefore $h$ uniformly by $0.002$ (so $h=10^{-4}$).

The effective resolution $\Delta x / R$, at the same radii, is not trivial to estimate in these numerical methods. We therefore consider three definitions, all specifically designed to characterize the fixed-grid Eulerian codes used in S25 and G25 (though the first definition below would need modification for Eulerian AMR codes, where vertical refinement could be much finer than uniform coverage). First, an ``equivalent Cartesian'' spatial resolution  $\langle \Delta x \rangle^{\rm C} \equiv ({\rm Volume}/N)^{1/3}$, counting the total number of particles and volume in a cylindrical annulus $R-\Delta R/2 < R < R + \Delta R/2$ with $\Delta R=R/2$ and $|z| < R$; this reflects the uniform resolution maintained across the entire domain in static-mesh codes. Second, the median actual resolution in the midplane region $\langle x \rangle^{\rm mid}$, defined by the median $\Delta x \equiv (\Delta m/\rho)^{1/3}$ of all cells within the small cylindrical ring around the midplane with $-H_{\rho}/2 < z < H_{\rho}/2$. Third, the resolution at which vertical structure in the Lagrangian code would become ill-defined, i.e.\ the disk would be effectively ``one cell thick,'' defined as in \citet{hopkins:superzoom.disk} by the cell separation which would be present if all cells within an annulus were placed in the midplane, $\langle\Delta x \rangle^{\rm 2d} \equiv ({\rm Area}/N)^{1/2}$ in terms of the annular area and $N$ used for $\langle \Delta x \rangle^{\rm C}$.

Figs.~\ref{fig:timeseries}-\ref{fig:profiles} show the numerical results from these tests. 
Although for Eulerian methods with poor resolution $\Delta x \gtrsim H_{\rm thermal}$, the disk thickness and $B_{\phi}$ are maintained, in all of our Lagrangian tests, we see a similar ``collapse'' begin as seen only at high resolution in Eulerian methods. Specifically, we see a decrease in time of the midplane density scale-height as a steep density ``spike'' forms at $|z| \sim h$, with a thicker, $\beta \ll 1$ envelope which contains an $\mathcal{O}(1)$ fraction of the total disk mass, and a ``dimple'' or decrease in the central magnetic pressure. The converged ($h=0.05$) Eulerian and Lagrangian solutions agree that collapse proceeds until $\beta \sim 1$ (density scale-height $\sim h$) in the midplane. However the low-resolution behavior differs: in Eulerian tests in S25 \&\ G25, no collapse or flux loss appears if $\langle \Delta x \rangle^{\rm C} \gtrsim H_{\rm thermal}$. In Lagrangian methods, the ``beginning'' of collapse appears to be essentially resolution-independent, even for $\langle \Delta x \rangle^{\rm C} \gg H_{\rm thermal}$. This is most obvious in the runs with extremely low-$\beta$ ($h=10^{-4}$). As we approach the cold plasma limit ($\beta\rightarrow 0$, $h\rightarrow 0$), the Lagrangian methods cannot resolve the $\beta \sim 1$ scale height $h$, so the code simply collapses to its effective resolution limit, $\sim \langle \Delta x \rangle^{\rm 2d}$, and so gets denser as we increase the resolution (approaching the converged solution). We see broadly similar results for MFM and MFV and runs with/without CG-MHD. 

There are some subtleties that may differ from G25 in the converged solutions: we see (as they noted) a strong association between overdense rings and spiral structures and vertical collapse, but the causal relationship remains unclear. As these structures move ``through'' a given annulus, the degree of collapse can vary azimuthally, and sometimes warps excited by grid noise from the collapsing material can ``stir up'' some regions, leading to the asymmetries visible in vertical pressure profiles (e.g., Fig.~\ref{fig:profiles}). Such azimuthal variations in pressure are natural consequences of the nonlinear dynamics during collapse and do not affect our primary conclusions about the resolution-dependent behavior between Lagrangian and Eulerian methods.

\section{Interpretation and Relation to Other Lagrangian-Eulerian Differences}
\label{sec:lagrangian}

There are some potentially simple interpretations of the difference at low resolution between Eulerian and Lagrangian methods in the G25 problem. In Eulerian fixed-mesh codes, $\Delta x$ at the midplane is fixed, and it is impossible to represent vertical structure/gradients in any field ($\rho$, ${\bf B}$, ${\bf v}$) below some multiple of $\Delta x$. Per S25/G25, when $\Delta x$ is too large, this makes it impossible for the Parker dynamo as described in \citet{johansen.levin:2008.high.mdot.magnetized.disks,gaburov:2012.public.moving.mesh.code} and S25 to shut itself off, and therefore prevents Parker modes from removing magnetic flux from scales $|z| \lesssim \Delta x$. In Lagrangian codes, there are two important qualitative differences. First, for a nominal Cartesian-equivalent $\langle \Delta x \rangle^{\rm C}$, the \textit{actual} vertical resolution $\langle \Delta x \rangle^{\rm mid}$ of the simulation around the midplane can be vastly superior, as by definition more gas will be concentrated towards the midplane, and any effect which attempts to initiate an increase in the midplane density (e.g.\ the early states of collapse) will lead to even better spatial resolution in the midplane. Indeed this one of the major motivations for Lagrangian methods in disk problems, in the first place. Second, in Lagrangian methods, cells \textit{move}, so it is always possible in principle for modes to carry magnetic flux away from the midplane with said cells, even in the limit when all cells have piled up in the midplane such that vertical gradients become numerically ill-defined ($\langle\Delta x \rangle^{\rm mid} \rightarrow \langle\Delta x \rangle^{\rm 2d}$). This explains why even taking $h \rightarrow 0$, we still see collapse (just down to the single-cell resolution limit, rather than $h$). 

This is analogous to other, better-studied differences between Eulerian and Lagrangian codes on collapse problems, perhaps most notably the ``artificial fragmentation'' problem. In simple test problems of seeded Jeans fragmentation in a homogeneous self-gravitating medium, as well as more complex problems, it is well-established \citep[see e.g.][]{truelove:1997.jeans.condition,truelove:1998.gmc.frag,banerjee:2004.protostellar.sphere.collapse} that in Eulerian (static or adaptive mesh refinement) methods, if $\Delta x > \lambda_{J} \equiv c_{s} / \sqrt{G \rho}$, numerically spurious modes begin to grow, and the mass/size distribution of fragments can be corrupted (causing ``over-fragmentation'') on scales much larger than $\Delta x$. But it is equally well-established that this error does not appear in Lagrangian methods (including SPH, MFM, and MFV; \citealt{hubber:2006.resolution.requirements.fragmentation.sph,chiaki:2015.particle.splitting.truelove.criterion.not.correct.for.lagrangian.codes,manuel:2016.no.effects.from.truelove.criterion,guszejnov:2018.isothermal.nocutoff,yamamoto:2021.no.artificial.fragmentation.in.mfm.mfv.sph}). So long as one uses a (required and always true in modern codes like GIZMO) consistent definition of force softening and hydrodynamic resolution, then in Lagrangian codes there is no over-fragmentation: instead the error if the Jeans length is unresolved ($\Delta x > \lambda_{J}$, equivalent to saying the Jeans mass is unresolved $\Delta m > M_{J}$) is simply the obvious one that collapse of fragments approaching the resolution limit (near single-cell) is slowed down, and obviously no fragments can form below the resolution $\Delta m$. Similar results hold for magnetized Jeans collapse \citep{myers:2013.trulove.with.mhd,guszejnov:2020.mhd.turb.isothermal.imf.cannot.solve} and disk fragmentation \citep{robertson:2008.molecular.sflaw,deng:gravito.turb.frag.convergence.gizmo.methods,forgan:2017.mhd.gravitoturb.sims,deng:2021.magnetic.disk.frag.in.gizmo.small.planetesimals,xu:2025.grav.instability.protostellar.disks.gravitoturb.stochastic.fragmentation.as.predicted,xu:2024.ppd.fragmentation.is.stochastic.as.predicted.hopkins.christiansen.mapped.with.rhd}. The generally-accepted interpretation of this is similar to our argument above: in Lagrangian codes, the fact that the cell size and mesh-generating point positions are shrinking and moving continuously with the fluid in the compressive modes means that the salient numerical error terms cannot propagate to larger scales.

On top of these, in different codes, the thermal speed can play a non-trivial role in numerical dissipation/viscosity, even in problems approaching the cold plasma limit. In cold shearing disks studies \citep[][and references therein]{imaeda:2002.sph.shear.flow.tests,hopkins:gizmo,schaal:2015.dg.amr.tests.including.keplerian.disk,duffell:2016.disco.code.disks,zier:2022.cold.shear.flow.sims.moving.meshes.error.control}, it is known that static Cartesian mesh methods like those used in G25 suffer from certain numerical errors and instabilities which are suppressed when the thermal scale-height is well-resolved, If these effects are important, we would expect a difference in this problem in Eulerian codes using Cartesian vs.\ well-aligned curvilinear meshes, or shearing boxes, so the similarity of S25 and G25 suggests this may not be a dominant effect.

\section{Discussion: Implications for Other Toroidally-Magnetized Disk Simulations}
\label{sec:discuss}

Strongly toroidally-magnetized disks have now been seen with a number of different codes using different numerical methods, including: 
FVMHD3D \citep{gaburov:2012.public.moving.mesh.code}, a Voronoi-tesselation based moving-mesh finite-volume scheme, in simulations of molecular cloud collisions with a black hole \citep{gaburov:2012.public.moving.mesh.code}; 
ATHENA-K \citep{stone:2024.athenaK}, with Eulerian static-but-nested Cartesian meshes, in ideal-MHD plus cooling simulations of low-level Bondi accretion of hot gas from galaxy to supermassive black holes \citep{guo:2024.fluxfrozen.disks.lowmdot.ellipticals}; 
ATHENA-K with similar nested meshes in ideal-MHD plus cooling simulations of collapse of turbulent clouds onto binary supermassive black holes \citep{wang:2025.hypermagnetized.circumbinary.disk.flux.frozen.cavity.to.pc.scales}; 
GIZMO \citep{hopkins:cg.mhd.gizmo}, with Lagrangian meshless-finite-mass (MFM) and meshless-finite-volume (MFV) methods, in simulations of high-redshift starburst galactic flows to supermassive black holes as quasars \citep{hopkins:superzoom.disk,hopkins:superzoom.overview,hopkins:superzoom.imf,hopkins:superzoom.agn.disks.to.isco.with.gizmo.rad.thermochemical.properties.nlte.multiphase.resolution.studies}; 
GIZMO using the MFM solver but a different numerical MHD method \citep{hopkins:mhd.gizmo}, in simulations of molecular-cloud and star-cluster accretion onto intermediate-mass black holes \citep{shi:2024.seed.to.smbh.case.study.subcluster.merging.pairing.fluxfrozen.disk,shi:2024.imbh.growth.feedback.survey}; 
ENZO \citep{oshea:2004.enzo.introduction}, with Eulerian Cartesian adaptive mesh refinement (AMR), in ideal-MHD simulations of gas inflows around first-stars \citep{luo:2024.magnetically.dominated.disk.like.our.zoomins.zoomin.on.first.supermassive.star.situation}; 
and
H-AMR \citep{liska:2022.hamr}, a general-relativistic MHD code using a static-mesh-refinement (SMR) spherical coordinate grid, in simulations \citep{kaaz:2024.hamr.forged.fire.zoom.to.grmhd.magnetized.disks} of inflows to horizon scales using an initial condition from \citet{hopkins:superzoom.disk}. 

This raises important questions. Could this imply that the persistence of toroidally-magnetized disks in the simulations owes to lack of resolution? 
This seems unlikely, as the simulations in \citet{shi:2024.seed.to.smbh.case.study.subcluster.merging.pairing.fluxfrozen.disk,shi:2024.imbh.growth.feedback.survey,hopkins:superzoom.disk,hopkins:superzoom.agn.disks.to.isco.with.gizmo.rad.thermochemical.properties.nlte.multiphase.resolution.studies} 
had $\Delta x \sim 0.01-0.1\,H_{\rm thermal}$ for the diffuse, volume-filling midplane gas, although their disks were multi-phase so they contain cold gas clouds with much smaller $H_{\rm thermal}$. 
The highest-resolution run in \citet{gaburov:2012.public.moving.mesh.code}, which was isothermal, had $\Delta x \approx 0.33\,H_{\rm thermal}$. 
\citet{hopkins:superzoom.agn.disks.to.isco.with.gizmo.rad.thermochemical.properties.nlte.multiphase.resolution.studies} re-ran multiple versions of their disks enforcing a temperature floor such that $\Delta x \lesssim 0.1\,H_{\rm thermal}$ for all gas. 
\citet{kaaz:2024.hamr.forged.fire.zoom.to.grmhd.magnetized.disks} ran for extended durations with an simplified effective equation of state (for radiation+gas) such that there was a well-defined $\Delta x \sim 0.02-0.05\,H_{\rm thermal}$ in the toroidally-magnetized zone. And \citet{luo:2024.magnetically.dominated.disk.like.our.zoomins.zoomin.on.first.supermassive.star.situation} had $\Delta x \sim 0.03\,H_{\rm thermal}$ in their outer disk, though these disks also had strong poloidal fields.
All of these meet the G25 ``high-resolution'' criterion. 
Moreover, our tests indicate that the Lagrangian simulations, including those from GIZMO and FVMHD3D (and perhaps ENZO, which is quasi-Lagrangian with AMR and so might exhibit some Lagrangian characteristics, but this is not obvious), can reproduce the initial vertical collapse and toroidal flux loss of G25 even at much worse resolution (they would just collapse to $\sim 1$ cell thick, which clearly does not happen in those simulations). 
Plus, all these simulations contain some non-negligible (if not dominant) poloidal field, which G25 found aided in maintaining midplane $\beta \ll 1$ at high resolution. 
So this very clearly argues that the sustained toroidal fields in those simulations are not a numerical artifact. 

Alternatively, could this imply that the ``collapse'' seen in S25 and G25 is a numerical artifact? Those simulations used static, Cartesian, Eulerian grids, which are well-known to be much more numerically diffusive/resistive in thin-disk problems \citep[see e.g.][]{hahn:2010.disk.gal.orientations.ramses,duffell:2011.TESS,duffell:2012.disco.method.protoplanetary.disk,duffell:2024.santa.barbara.binary.disk.code.comparison.gizmo.tests.comparable.ideal.performance,mocz:2014.galerkin.arepo,munoz:2014.disk.planet.interaction.sims,hopkins:gizmo,seo:2019.gas.gizmo.bar.formation.sims,deng:2019.mri.turb.sims.gizmo.methods,deng:2020.global.magnetized.protoplanetary.disk.sims.gravito.turb.leads.to.large.B.saturation.vs.mri,deng:2020.parametric.instab.free.disks,deng:2022.warped.disk.dynamics.need.mfm.mfv} compared to either Eulerian spherical/cylindrical coordinate grids (like in \citealt{kaaz:2024.hamr.forged.fire.zoom.to.grmhd.magnetized.disks}) or Lagrangian grids (like in \citealt{gaburov:2012.public.moving.mesh.code,hopkins:superzoom.disk,shi:2024.seed.to.smbh.case.study.subcluster.merging.pairing.fluxfrozen.disk}). 
S25, for example, saw the midplane $\beta$ increase from an initially low value to $\beta \sim 1$ as their resolution improved from $\Delta x \approx  H_{\rm thermal}$ to $\Delta x \approx 0.4\, H_{\rm thermal}$, but then saw the midplane $\beta$ gradually decrease again steadily with further resolution improvements up to their highest resolution $\Delta x \approx 0.07\,H_{\rm thermal}$, without an obvious indication of convergence. 
%The only other Cartesian-mesh example of the simulations above is that in \citet{luo:2024.magnetically.dominated.disk.like.our.zoomins.zoomin.on.first.supermassive.star.situation}, where the maintenance of a strong toroidal midplane magnetic field could potentially be explained by the presence of a comparably strong ($|B_{z}| \sim |B_{\phi}|$) vertical field, as both S25 and G25 saw a similar effect in their idealized tests. 
But our results imply that collapse on this test problem is not purely an artifact of the Eulerian method, since we see the same collapse in Lagrangian simulations with different methods, although we cannot definitively answer the question of convergence in this test problem either (as we do not replicate S25). 

Instead, our results argue that that the difference between the idealized, symmetric, non-turbulent, strictly isothermal ideal-MHD simulation tests in S25, G25, and herein, and the multi-scale, multi-physics simulations above, likely owes to some real physics or initial/boundary conditions. 
There is no shortage of plausible candidates, as discussed in e.g.\ \citet{hopkins:superzoom.disk}: the simulations in \citet{hopkins:superzoom.disk,hopkins:superzoom.overview,hopkins:superzoom.imf,hopkins:superzoom.agn.disks.to.isco.with.gizmo.rad.thermochemical.properties.nlte.multiphase.resolution.studies,kaaz:2024.hamr.forged.fire.zoom.to.grmhd.magnetized.disks,luo:2024.magnetically.dominated.disk.like.our.zoomins.zoomin.on.first.supermassive.star.situation,shi:2024.seed.to.smbh.case.study.subcluster.merging.pairing.fluxfrozen.disk,shi:2024.imbh.growth.feedback.survey,wang:2025.hypermagnetized.circumbinary.disk.flux.frozen.cavity.to.pc.scales} not only include much more complex, multi-phase, turbulent, highly dynamic, inhomogeneous, non-spherical boundary/initial conditions, but they also include physics like self-gravity, multi-group radiation-hydrodynamics and radiation pressure, multi-phase gas thermo-chemistry with complex equations-of-state, general relativity, star formation and stellar feedback, collisionless (stars+dark matter) plus gas components, highly asymmetric disks with strong global modes and spiral arms, etc. So the challenge is isolating any single piece of physics that could be most important. A hint might come from the more simplified ideal-MHD, analytic-gravity simulations in \citet{gaburov:2012.public.moving.mesh.code} which also adopted a simple locally-isothermal equation-of-state, but whose initial condition allowed for a more turbulent and asymmetric (eccentric with strong global modes) disk, with a much larger accretion/inflow (and therefore toroidal flux \textit{replenishment}) rate than G25. Alternatively the simulations in \citet{wang:2025.hypermagnetized.circumbinary.disk.flux.frozen.cavity.to.pc.scales} adopt a  turbulent initial condition, with simplified but multi-phase gas cooling, and an asymmetric time-dependent (binary) analytic background. And as noted above all these more complex simulations included non-trivial poloidal field structure as well.

\section{Summary}
\label{sec:summary}

We reproduce the test problem from G25 -- collapse of a homogeneous, uniformly rotating and magnetized sphere in isothermal ideal-MHD with Keplerian gravity -- with multiple Lagrangian methods. We show that Lagrangian methods reproduce the high-resolution behavior of the G25 Eulerian tests, wherein the midplane becomes denser with $\beta\sim1$. S25 and G25 showed that in Eulerian codes, seeing almost any flux loss or density increase requires resolving the thermal scale-length of the disk, $\Delta x < H_{\rm thermal} = c_{s}/\Omega$. But in Lagrangian methods, the initial flux loss and density increase is recovered at all resolution levels even when the thermal scale-length is arbitrarily poorly-resolved. In the unresolved limit, collapse simply proceeds until the midplane reaches single-cell thickness (reaching ``as close as possible'' to the high-resolution solution). We argue this owes to the fundamentally different behavior of Lagrangian methods in their ability to follow vertical motions and structure in the limit where the thermal scale-length is unresolved. This is analogous to other well-studied differences between Lagrangian and Eulerian codes, notably their very different behaviors in Jeans fragmentation.

This argues that the \textit{lack} of such collapse -- i.e.\ sustained, strong toroidal fields -- seen in the multi-physics simulations which motivated S25 and G25, is numerically robust. In other words, in addition to the fact that the simulations and numerical tests in \citet{hopkins:superzoom.disk,hopkins:superzoom.agn.disks.to.isco.with.gizmo.rad.thermochemical.properties.nlte.multiphase.resolution.studies,kaaz:2024.hamr.forged.fire.zoom.to.grmhd.magnetized.disks,luo:2024.magnetically.dominated.disk.like.our.zoomins.zoomin.on.first.supermassive.star.situation} already satisfy the resolution criterion argued for in G25, the fact that the simulations in \citet{hopkins:superzoom.disk,hopkins:superzoom.agn.disks.to.isco.with.gizmo.rad.thermochemical.properties.nlte.multiphase.resolution.studies,shi:2024.seed.to.smbh.case.study.subcluster.merging.pairing.fluxfrozen.disk,shi:2024.imbh.growth.feedback.survey} use the same Lagrangian methods tested here (at much higher resolution), which reproduce collapse in the G25 test problem even at much worse resolution, argues that the lack of such collapse in their multi-physics simulations is not a resolution artifact. A much more likely hypothesis is that some other physics or boundary/initial conditions explains the difference. Future work is clearly called for to identify those key drivers.

\begin{acknowledgements}
We thank Minghao Guo, Eliot Quataert and Jono Squire for helpful discussions. Support for PFH was provided by a Simons Investigator Grant. Numerical calculations were run on NSF TACC allocation AST21010.
\end{acknowledgements}

\bibliographystyle{mn2e}
%\bibliography{/Users/phopkins/Dropbox/Public/ms}
\bibliography{ms_extracted}

\end{document}